\documentclass[hyper,a4paper,12pt]{JHEP3}

\usepackage{graphicx}
\usepackage{epsfig}
\usepackage{amsmath}
\usepackage{amssymb}
\usepackage{amsthm}

\newcommand{\non}{\nonumber \\}

\newcommand{\alp}{\alpha}     
     \newcommand{\del}{\delta}
\newcommand{\eps}{\epsilon}   
      \renewcommand{\th}{\theta}

   \newcommand{\sig}{\sigma}
 
   \newcommand{\ome}{\omega}

\newcommand{\Sig}{\Sigma}     
\newcommand{\Ome}{\Omega}

    \newcommand{\cN}{{\cal N}}
\newcommand{\cO}{{\cal O}}

\newcommand{\RR}{\mathbb{R}}

\newcommand{\pa}{\partial}

\newcommand{\rar}{\rightarrow}

\newcommand{\gsim}{ \lower .75ex \hbox{$\sim$} \llap{\raise .27ex \hbox{$>$}} }
\newcommand{\lsim}{ \lower .75ex \hbox{$\sim$} \llap{\raise .27ex \hbox{$<$}} }

\title{Anisotropic Drag Force from 4D Kerr-AdS Black Holes}
\author{A. Nata Atmaja${}^{1,2}$ and K. Schalm${}^{1,2}$\\
${}^1$Institute Lorentz for Theoretical Physics\thanks{Permanent address.}\\
Niels Bohrweg 2\\
2333CA Leiden, the Netherlands\\
\\
${}^2$Instituut voor Theoretische Fysica\\
Valckenierstraat 65\\
1018XE Amsterdam, the Netherlands\\
\email{ardian@lorentz.leidenuniv.nl, kschalm@lorentz.leidenuniv.nl}
}

\abstract{Using AdS/CFT
we investigate the effect of angular-momentum-induced anisotropy on the instantaneous drag force of a heavy quark. The dual description is that of a string moving in the background of a Kerr-AdS black holes. The system exhibits the expected focussing of jets towards the impact parameter plane.
We put forward that we can use the connection between this focussing behavior and the angular momentum induced pressure gradient to extrapolate the pressure gradient correction to the drag force that can be used for transverse elliptic flow in realistic RHIC. The result is recognizable as a relativistic pressure gradient force.}

\preprint{ITFA-2010-24}
\begin{document}

\section{Introduction}
Heavy ion collisions at RHIC with energy around 200 GeV per nucleon are believed to produce sQGP (strongly coupled quark gluon plasma)~\cite{Shuryak:2006se}. In this region, perturbative QCD is no longer reliable and it is necessary to explore all non-perturbative calculational methods of QCD. One of the elegant and current developing techniques is the string/gauge duality in the form of the AdS/CFT correspondence~\cite{maldacena97,witten98-1,Gubser:1998bc}.
Although there is no known AdS dual to QCD , it is believed that 
finite temperature 
four-dimensional conformal field theory~\cite{witten98-2} shares many features with the sQGP phase of QCD as long as the temperature is above the deconfiment scale~\cite{Gubser:2006bz}. In particular collective properties of strongly coupled fluids such as the sQGP should be well described in an AdS dual description~\cite{Bhattacharyya:2007vs,Son:2007vk}.

An important feature of RHIC is the spatial anisotropy of the data correlated with non-zero impact parameter of the colliding ions. The spatial anisotropy translates into a non-zero component of the second Fourier harmonic of the particle distribution in the plane transverse to the collision.\footnote{The first Fourier component vanishes by symmetry.} This coefficient is known as elliptic flow and has been calculated using hydrodynamic evolution of the sQGP  in a stunning agreement with data~\cite{Kolb:2000sd,Kolb:2000fha}.\footnote{A more detailed discussion on this phenomena and others can be found in~\cite{Kolb:2003dz} and references therein.}
Up to now most AdS computations of sQGP have ignored spatial anisotropy: the black hole background is always the standard static isotropic AdS black hole. There is good reason to do so. Transport coefficients such as the viscosity are defined with respect to the isotropic perfect fluid and for other quantities the experimental indication that the system thermalizes rapidly to an almost perfect, i.e. isotropic, fluid means that anisotropy corrections are small. The exceptions are ``local temperature/pressure'' approximations as in Bhattacharyya et al~\cite{Bhattacharyya:2008xc} and Chesler et al~\cite{Chesler:2008hg} and shockwave collision set-ups \cite{Lin:2009pn,Gubser:2009sx,Avsar:2009xf,DuenasVidal:2010vi,Taliotis:2010pi}. Yet in practice isotropic observables are polluted by the
considerable spectator detritus of the heavy ion collision.  Due to
elliptic flow , the anisotropic component of any plasma is enhanced
w.r.t. average isotropic background. It is therefore the anisotropic
component that is in many cases experimentally the most accessible
avenue to study the collective properties of the plasma\footnote{There are anisotropies of observables which are not due to
collective effects.  Methods  exist to identify the "collective
anisotropy" and separate it from the anisotropies caused by e.g.
resonance decays, jets etc. For the second Fourier component these are
what we call "flow" and "non-flow". Such a separation is much more
difficult for isotropic parts of an observable. We thank Thomas
Peitzmann for explaining this to us.}.

In this paper we make a first step towards the study of anisotropic effects on jet-quenching from the string theory point of view. Jet-quenching
is a characteristic feature of the sQGP phase in RHIC. It signals the strong energy loss of a highly massive quark moving in hot charged plasma. In the frame work of AdS/CFT, a heavy quark is represented by a string suspended from the boundary of asymptotically AdS space into the interior~\cite{Maldacena:1998im,Rey:1998bq,Rey:1998ik}.
This set-up was proposed in the context of $\cN=4$ supersymmetric Yang-Mills theory at finite temperature~\cite{Herzog:2006gh,Gubser:2006bz} and has been explored in detail in \cite{Chesler:2007an,Chesler:2007sv} with a beautiful extension to the trailing wake of the heavy quark in the sQGP dual to the backreaction of string on the black-hole geometry. 
The way we shall introduce anisotropy in the system is to consider non-zero angular momentum. The advantage is that the dual description of this system is straightforward: one considers rotating black holes. The drawback is that the anisotropy primarily responsible for elliptic flow  is due to the asymmetric almond-shape overlap of the two non-central colliding nuclei rather than angular momentum \cite{Voloshin:2008dg}. As all non-central collisions, the total system does carry a significant amount of angular momentum, but most of that is carried away by spectator-nuclei not involved in the formation of the sQGP. At RHIC the angular momentum fraction of the total elliptic flow is thought to be less than 10\%, although it is expected to increase to 30\% at LHC (see~\cite{Becattini:2007sr,Abreu:2007kv} and the references therein). Clearly  
an AdS/CFT study of elliptic flow due to non-rotational anisotropy would be  more relevant. The problem is that the gravity set-up in this case is unclear. Non-rotational anisotropy dissipates fast as the system equilibrates and isotropizes, and this points to a time-dependent gravity dual, along the lines of \cite{Chesler:2008hg}\footnote{Alternatively one could consider 4+1 dimensional hairy black holes to break the anisotropy; we thank H. Ooguri for pointing this out.}. We shall nevertheless propose that studying anisotropic jet-quenching in a rotating plasma may contain meaningful information for transverse elliptic flow. A rotating relativistic fluid has a pressure gradient. If one can argue that certain components of the anisotropic drag force experienced in a rotating fluid are in effect solely due to this pressure gradient and if one can extract these components and their coefficient strengths, then one can put forward that the same corrections with the same strengths arise due to the transverse pressure gradients relevant for realistic elliptic flow. Doing so, one infers that the leading contribution to the drag force at strong coupling due to anisotropic pressure in the plasma equals a relativistic pressure gradient force, familiar from e.g. astrophysics
\begin{eqnarray}
  \label{eq:1}
  \Delta\vec{F}_{drag~due~to~anisotropy} =  -\left(3 m_{rest}\right) \frac{\nabla P}{sT} +\ldots
\end{eqnarray}
Here
 $s, T$ are the (nearly constant) entropy density and temperature of the plasma, and $m_{rest}$ is the rest mass of the heavy quark, which must be larger than any scale in the system.

For simplicity we shall consider only 3+1 dimensional AdS black holes dual to 2+1 dimensional thermal field theories. A technical complication is that a rotating relativistic fluid is only consistent on a sphere and we are forced to consider the global description of AdS blackholes. The resulting calculation is then very similar to the holographic drag force of charged quarks in a charged plasma \cite{Herzog:2006se,Caceres:2006dj,Herzog:2007kh,Atmaja:2010uu}. This holographic drag force is equivalent to strings moving in a $d=10$ spinning D3-brane background, whose near horizon limit is an AdS${}_5 \times S_5$ Kerr black hole with has rotation along the internal $S^5$ direction. In the calculation here the sphere is now the geometric sphere of the AdS boundary $\pa{\rm AdS} \simeq \RR \times S_3$ on which the dual gauge theory lives.  In Sec.2, we first recover the known drag force in the global description from strings with endpoints that move along great circles (geodesics on the boundary sphere). In Sec. 3, we introduce angular momentum by generalizing the background metric to a rotating 3+1 dimensional Kerr-AdS black hole. In the limit of small angular momentum and for zero initial velocities, we can compute the leading correction to drag force at an arbitrary initial location. The result exhibits the expected focussing in the impact-parameter plane. In the conclusion we show how the leading anisotropic correction to the drag force due to non-zero angular momentum implies the pressure gradient correction Eq. \eqref{eq:1}.

\section{Drag force on a string in a global 4D AdS black hole}

In global coordinates, the metric of four dimensional AdS-Schwarzschild is given by 
\begin{eqnarray}
\label{eq1}
 ds^2&=&-r^2h(r)dt^2+\frac{1}{r^2h(r)}dr^2+r^2\left(d\theta^2+\sin^2\theta d\phi^2\right),\\
 h(r)&=&l^2+\frac{1}{r^2}-\frac{2M}{r^3},\nonumber
\end{eqnarray}
where $M$ is proportional to the mass of the black hole and $l$ is the radius of curvature. The Hawking temperature of four dimensional AdS-Schwarzschild can be obtained in a simple way 
by demanding periodicity of Euclidean time such that we avoid a conical singularity at $r=r_H$. This gives $T_H=\frac{1}{4\pi}\left(\frac{1}{r_H}+3r_Hl^2\right)$, where $r_H$ is the radius of horizon defined as the zero locus $h(r_H)=0$~\cite{Hemming:2007yq} and can be written explicitly in terms of parameters $l$ and $M$: (Note that $M$ has dimensions of length and $l$ has dimensions of mass.)
\begin{align}  
r_H(l,M)=\frac{\left(9Ml^4+\sqrt{3l^6+81M^2l^8}\right)^{1/3}}{3^{2/3}l^2}-\frac{1}{3^{1/3}\left(9Ml^4+\sqrt{3l^6+81M^2l^8}\right)^{1/3}}.
\end{align}
The Poincare patch black hole that is usually used in AdS/CFT studies is obtained by rescaling $\{\theta,r,t,M\} \rar \{\eps\theta,r/\eps,\eps t,M/\eps^3\}$ and taking the limit $\eps\rar 0$.

A string in this background can be described by the Nambu-Goto action:
\begin{align}
\label{eq2}
 S=&-\frac{1}{2\pi\alpha'}\int{d\sigma^2\sqrt{-\mbox{det} g_{\alpha\beta}}}=\int{d\sigma^2\mathcal{L}},\notag\\
g_{\alpha\beta}\equiv& G_{\mu\nu}\partial_\alpha X^\mu\partial_\beta X^\nu,
\end{align}
with $\sigma^\alpha$ are coordinates of string worldsheet, $X^\mu=X^\mu(\sigma)$ are the embedding of string worldsheet in spacetime, and $G_{\mu\nu}$ is the spacetime metric (\ref{eq1}).
The equation of motions derived from (\ref{eq2}) are,
\begin{equation}
\label{eq3}
 \nabla_\alpha P^\alpha_\mu=0,\ \ \ \ P^\alpha_\mu\equiv-\frac{1}{2\pi\alpha'}G_{\mu\nu}\partial^\alpha X^\nu =- \frac{1}{2\pi\alp' \sqrt{-g}}\, \pi^{\alp}_{\mu}~,
\end{equation}
with $g=\mbox{det}g_{\alpha\beta}$. Here $ P^\alpha_\mu$ is the worldsheet current of spacetime momentum carried by the string and $\pi^\alpha_\mu$ equals the canonical worldsheet momentum,
\begin{align}
\pi_{\mu}^{\alp} = -(2\pi\alp')\frac{\del S}{\del \pa_{\alp}X^{\mu}}.
\end{align}
The total momentum charge in the direction $\mu$ carried by the string equals
\begin{eqnarray}
p_{\mu}= \int d\Sig_{\alp}\sqrt{-g}P^{\alp}_{\mu},
\end{eqnarray}
where $\Sig_{\alp}$ is a cross-sectional surface (a line) on the string worldsheet. Time independent forces are given by momentum-flow along the string across a time-like surface $d\Sig_{\sig^1}=d\sig^0 \sqrt{-g_{00}}\hat{n}_{\sig^1}$ \cite{Herzog:2006gh}. The proper-force on the string then equals 
\begin{eqnarray}
  \frac{\pa p_{\mu}}{\pa \sig^0}&=& \sqrt{-g} P^{\sig^1}_{\mu}
\end{eqnarray}
which in turn is equal to the canonical worldsheet-momentum
\begin{eqnarray}
  \label{eq:4}
  \frac{\pa p_{\mu}}{\pa \sig^0} &=& -\frac{1}{2\pi\alp'} \pi^{\sig^1}_{\mu} 
\end{eqnarray}
If the configuration is constant in time, the proper force $\frac{\pa p_{\mu}}{\pa \sig^0}$ does not depend on the location $\sig^1$ along the worldsheet, thanks to the equation of motion.
\begin{eqnarray}
 \label{eq:3}
  \frac{\pa}{\pa {\sig^1}} \frac{\pa p_{\mu}}{\pa \sig^0} = \frac{\pa}{\pa \sig^1} \sqrt{-g} P^{\sig^1}_{\mu} =-  \frac{\pa}{\pa \sig^0} \sqrt{-g} P^{\sig^0}_{\mu} \stackrel{static}{=}0
\end{eqnarray}

Using the physical gauge, $\sigma^\alpha=(t,r)$, we can write the action (\ref{eq2}) as follows
\begin{eqnarray}
\label{eq4} 
S&=&-\frac{1}{2\pi\alpha'}\int{d\sigma^2\sqrt{
 -g}}, \\\nonumber
 -g&=&
1+r^4h(r)\left(\theta'^2+\phi'^2\sin^2\theta\right)-\frac{1}{h(r)}\left(\dot{\theta}^2+\dot{\phi}^2\sin^2\theta\right)-r^4\sin^2\theta\left(\dot{\theta}\phi'-\theta'\dot{\phi}\right)^2,
\end{eqnarray}
with $\dot{\theta}\equiv{d\over dt}\th$, etc. and  $\th'\equiv{d\over dr}\th$. etc.

\subsection{Equatorial quark trajectories: Great circle at $\theta=\pi/2$}
\label{sec:equitorial great circle}

The boundary of the global AdS black hole metric is isomorphic to the sphere $S^2$. In the absence of any plasma, i.e. no black hole in the interior of AdS, we expect the heavy quark to follow the geodesic trajectory of a free particle on the sphere or great circles. To study how great circle motion is affected by the presence of plasma, we make a steady state ansatz, where the quark is dragged along the geodesic with constant velocity throught the medium. The string in the interior of AdS will then curve to spread the resulting tension along the worldsheet. From its resulting equilibrium shape we can compute the drag following \cite{Gubser:2006bz,Herzog:2006gh}. 

The simplest steady state solution is when we consider motion in the equatorial plane $\theta=\pi/2$ only. The rotational symmetry of AdS-Schwarzschild will guarantee that the full string curves inside this plane and we can make an ansatz for $\phi(r)$ as 
\begin{equation}
\label{eq7}
 \phi(t,r)=\omega t+\eta(r),
\end{equation}
with $\omega$ the angular velocity. Substituting this ansatz (\ref{eq7}) into the action (\ref{eq4}), we obtain an effective action for $\eta(r)$. Its momentum conjugate $\pi_{\phi}^r$ is a constant of the motion (angular momentum)
\begin{eqnarray} \pi^r_\phi&=&-2\pi\alpha'\frac{\partial\mathcal{L}}{\partial\eta'(r)}=\frac{r^4h(r)\eta'(r)}{\sqrt{1+r^4h(r)\eta'(r)^2-{\omega^2\over h(r)}}},
\end{eqnarray}
In terms of $\pi_{\phi}^r$ the equation of motion is
\begin{eqnarray}
 \label{eq8} \eta'(r)&=&\frac{\pi^r_\phi}{r^4h(r)}\sqrt{\frac{h(r)-\omega^2}{h(r)-\frac{(\pi^r_\phi)^2}{r^4}}},
\end{eqnarray}
In order to have a sensible solution, we have to require that (\ref{eq8}) must be real everywhere. This requirement gives us a condition on the argument of the square root
\begin{equation}
\label{eq9}
 \frac{h(r)-\ome^2}{h(r)r^4-(\pi^r_\phi)^2} \geq 0
\end{equation}
Because (\ref{eq8}) ought to be real everywhere for $r_H\leq r<\infty$, the only possible choice is to take the constant $(\pi^r_\phi)^2=\omega^2r_{SchErg,\ome}^4$, with $r_{SchErg,\ome}$ defined to be the point where $h(r_{SchErg,\ome})=\ome^2$. Then the numerator and denominator in (\ref{eq9}) change their sign at the same point at $r=r_{SchErg,\ome}$. (Note that the reality requirement \eqref{eq9} at the boundary $r\to\infty$ tells us that the velocity of the particle cannot exceed the speed of light $1\geq \omega^2/l^2$.\footnote{The extra factor of $l$ follows from the fact that the boundary metric corresponding to \eqref{eq1} equals \begin{align}
\label{metric boundary}
  ds^2_B =-dt^2+ {1\over l^2}\left(d\theta^2+ \sin^2\theta d\phi^2\right).
\end{align}})

The exact solution for equation (\ref{eq8}) is quite difficult to find but to compute the drag force it is enough to use the equation (\ref{eq8}). To compute the flow of momentum $dp_{\phi}$ down the string, we use
\begin{eqnarray}
\label{eq12}
\Delta  P_\phi&=&\int d\Sig_{\alp}P^{\alp}_{\phi}.
\end{eqnarray}
In this static configuration all momentum flow is radial. Thus the total momentum reduces to 
\begin{eqnarray}
\Delta P_{\phi}
&=&\int_\mathcal{I}dt\sqrt{-g}P^r_\phi=\frac{dp_\phi}{dt}\Delta t,
\end{eqnarray}
with $\mathcal{I}$ is some time interval of length $\Delta t$. Thus the drag force in the $\phi$ direction is given by
\begin{eqnarray}
\label{eq13}
 \frac{dp_\phi}{dt}=\sqrt{-g}P^r_\phi=-\frac{\pi^r_{\phi}}{2\pi\alpha'},
\end{eqnarray}
where the negative value implies that it is the drag force. 
Explicitly one can recognize in $\pi_{\phi}^r=\ome r_{SchErg,\ome}^2$ the centripetal acceleration at the ``ergosphere'' defined by $h(r_{SchErg,\ome})=\ome^2$.

\subsection{General great circle quark trajectories}
In general, the great circle can be in an arbitrary plane. By symmetry, however, the total force should be same for any great circle and be concentrated in the cross-sectional plane that defines the great circle when one views the $S^2$ as embedded in $S^3$. This symmetry will clearly be broken, however, when we introduce angular momentum and we therefore need to be able to describe general great circles that are not equatorial.  We can do so using an embedding in an auxiliary $S^3$
\begin{align}
 x=\sin\theta \cos\phi,\qquad\qquad y=\sin\theta \sin\phi,\qquad\qquad z=\cos\theta.\label{coord trans}
\end{align}
and considering the constrained Nambu-Goto action
\begin{eqnarray}
\label{eq26}
S_{cNG}&=&-\frac{1}{2\pi\alpha'}\int d\sigma^2\sqrt{-g}\left[1+ \frac{\lambda^2}{2}(x^ix^i-1)\right],\nonumber\\
-g&=&r^4\left((\dot{x}^ix'^i)^2-(\dot{x}^i\dot{x}^i-h(r))\left(x'^ix'^i+ \frac{1}{r^4 h(r)}\right)\right),
\end{eqnarray}
where $x^i\equiv(x,y,z)$ and $\lambda$ is a Lagrange multiplier. The equations of motion for this action are given by a constraint equation $x^ix^i=1$ and
\begin{eqnarray}
\label{eq27}
 \lambda^2x^i\sqrt{-g}&-&\frac{\partial}{\partial t}\left(x'^i\frac{r^4\dot{x}^jx'^j}{\sqrt{-g}}-\dot{x}^i\frac{\left(r^4x'^jx'^j+\frac{1}{h(r)}\right)}{\sqrt{-g}}\right)\notag\\
&-&\frac{\partial}{\partial r}\left(\dot{x}^i\frac{r^4\dot{x}^jx'^j}{\sqrt{-g}}-x'^i\frac{r^4(\dot{x}^j\dot{x}^j-h(r))}{\sqrt{-g}}\right)=0.
\end{eqnarray}
Notice that if we substitute the constraint equation back to the action (\ref{eq26}) then we get back the action (\ref{eq2}).

\subsubsection{The boundary geodesic}
The boundary geodesic followed by the endpoint of the string follows from considering the constrained Nambu-Goto action at a fixed $r$ for $r \gg 1/l$. Making a radial independent ansatz, $x^i=x^i(t)$, the equation of the boundary geodesic is
\begin{align}
\label{eq28}
 \lambda^2x^i\sqrt{-g}+\frac{\partial}{\partial t}\left(\frac{\dot{x}^i}{h(r)\sqrt{-g}}\right)=0,
\end{align}
where $-g={1\over h}(h-\dot{x}^i\dot{x}^i)$. Multiplying with $x^i$ and using the constraint $x^i\dot{x}^i=0$, we obtain
\begin{align}
\label{eq29}
\dot{x}^i\dot{x}^i={\lambda^2 h(r)\over 1+ \lambda^2}.
\end{align}
A natural solution would be to assume $\lambda$ is constant, in which case the equations of motion simplify to
\begin{align}
\ddot{x}^i=-{\lambda^2 h(r)\over 1+\lambda^2}x^i.
\end{align}
with the general solution 
\begin{equation}
\label{eq30}
 x^i(t)=a^i \sin\left(\lambda\sqrt{h(r)\over 1+\lambda^2}t\right)+b^i \cos\left(\lambda\sqrt{h(r)\over 1+\lambda^2}t\right),
\end{equation}
where $a^i$ and $b^i$ are constants. The constraint requires that these constants obey
\begin{equation}
 a^ia^i=b^ib^i=1,\qquad\qquad\qquad a^ib^i=0.
\end{equation} 
These solutions are the general great circle solutions in the plane spanned by $\vec{a}$ and $\vec{b}$. However, these solutions have an angular velocity which depends on $r$ and therefore they are not consistent with the ansatz of radial independence.

It is, however, straightforward to verify that the constant angular velocity $v$ configuration
\begin{equation}
\label{eq31}
 x(t)^i=a^i \sin\left(v t\right)+b^i \cos\left(v t\right).
\end{equation}
is also a solution, if $\lambda=\lambda(r)$ depends on the radial coordinate,
as
\begin{align}
\lambda(r)^2={v^2\over h(r)-v^2}.
\end{align}
Clearly this solution makes no sense deep in the AdS interior where $h(r)\ll v^2$, but for the boundary geodesic this is no issue. At the same time this is the  ``sign alternation'' we previously found for the string whose endpoint is the equatorial great circle. We therefore will take this solution as our starting point. 

(An alternate interpretation of this solution is as a moving straight string \cite{Herzog:2006gh}.)

\subsubsection{Drag Force from curved string solution}
Motivated by equatorial solution (\ref{eq7}) and radial independent general great circle solutions (\ref{eq31}), we take as ansatz for the curved string solution
\begin{equation}
\label{eq32}
 x^i(t,r)^i=a^i\sin(v t+c(r))+b^i\cos(v t+c(r)),
\end{equation}
with $v$ is a constant. Using the constraint equation, as we did in the radial independent ansatz, the equations of motion now become
\begin{align}
\label{eq33}
\sqrt{-g}{\partial\over \partial r}\left(r^4 h\ x^i\over \sqrt{-g}\right)=\left({\lambda^2\over 1+\lambda^2}-{v^2\over h}\right)x^i.
\end{align}
Substituting the ansatz (\ref{eq32}) into the induced metric, its determinant equals
\begin{align}
-g=1 + r^4 h(r)\ c'(r)^2-{v^2\over h(r)}.
\end{align}
Comparing this with the equatorial solution we can identify $c(r)=\eta(r)$ and $v=\omega$. Using the same way to get \eqref{eq29}, we obtain
\begin{align}
\lambda^2={v^2-r^4h^2c'^2 \over h+r^4h^2c'^2-v^2}
\end{align}
and therefore the determinant $-g$ can also be seen to equal $-g={1\over 1+\lambda^2}$. Using these two expressions of $-g$, one can check that at equatorial we get back the equatorial solution (\ref{eq7}). Now, the equations of motion reduce to 
\begin{align}
{\partial\over\partial r}\left(r^4 h\ c'(r)\over\sqrt{-g}\right)=0
\end{align}
which is the same equation for $\eta(r)$ as in the equatorial case and have the consistency condition $\pi_c^r= \ome r_{SchErg,\ome}^2$ where $r_{SchErg,\ome}$ is defined as $h(r_{SchErg,\ome})=v^2=\ome^2$.
Furthermore one can also find that $\lambda$ equals
\begin{align}
\lambda(r)^2={\ome^2\over r^4}{r^4-r^4_{SchErg,\ome}\over h(r)-\ome^2},
\end{align}
The definition of $r_{SchErg,\ome}$ guarantees that $\lambda(r)$ is a positive definite function.

The advantage of the constrained Nambu-Goto action (\ref{eq26}) is that the use of $x,y,z$ coordinates makes the $SO(3)$ rotation  symmetry manifest. As before, the associated angular momentum currents in the $r$ direction yield the drag force or rather drag torques
\begin{eqnarray}
\label{eq35}
\frac{dL^i}{dt}&=&-\frac{1}{2\pi\alp'}J^i_r\non
&=&-\frac{r^4}{2\pi\alp'\sqrt{-g}}\left(\dot{x}^jx'^j\varepsilon_{imn}x^m\dot{x}^n-(\dot{x}^j\dot{x}^j-h(r))\varepsilon_{imn}x^mx'^n\right),
\end{eqnarray}
where $\varepsilon_{imn}$ is a totally antisymmetric tensor, with $\varepsilon_{123}= 1$.
Substituting the ansatz and the constraint, we find 
\begin{equation}
\label{eq36}
\frac{dL^i}{dt}=-\frac{r^4 h(r)}{2\pi\alpha'}\frac{\varepsilon_{ijk}x^jx'^k}{\sqrt{-g}}=-\frac{r^4 h(r)c'(r)}{2\pi\alpha'\sqrt{-g}}\varepsilon_{ijk}b^ja^k=-\frac{r^4h(r)c'(r)}{2\pi\alpha'\sqrt{-g}}n^i
\end{equation}
where $n^i\equiv(n_x,n_y,n_z)$ is the normal vector to the cross-sectional plane defined by great circle, normalized to unity. Thus the force/torque is in the same or rather opposite direction of the motion. Furthermore, the equations of motion imply that these angular momentum currents of torques are constants. The norm of total angular momentum current or total torque is readily seen to be equal to the norm of equatorial drag force (\ref{eq13}),
\begin{eqnarray}
J^2_r\equiv J^i_rJ^i_r=\frac{dL^i}{dt}\frac{dL^i}{dt}=\left(\frac{dp^{(equatorial)}_\phi}{dt}\right)^2.
\label{eq:12}
\end{eqnarray}
This shows that also the square of total drag force is indeed the same for any great circle motion.\footnote{For nonequatorial motion, the RHS of (\ref{eq:12}) will contain nonzero $dp_\theta/dt$. Individually $dp_\theta/dt$ is not conserved, but the square of total drag force with the solution (\ref{eq32}) is a constant as given by the LHS of (\ref{eq:12}).}

\medskip

The drag force of a string moving in the background of 4D AdS-Schwarzschild is thus a constant related to the momentum of a particle represented by the end of the string at the boundary. To illustrate the frictional nature of the force
note that in the non-relativistic limit, the drag force (for equatorial motion) can be written as 
\begin{equation}
 \frac{dp_\phi}{dt}=-\frac{l^2}{2\pi\alpha'}\frac{p_\phi}{m}r_H^2+{\cal O}(\omega)
\end{equation}
with $p_{\phi}=m{\ome/l^2}$ 
the angular momentum. Taking $r_H=\frac{ 4\pi  T  + \sqrt{(4\pi T)^2-12 l^2}}{6l^2}$ as the plasma description corresponds to the solution with $r_H$ largest \cite{Bhattacharyya:2007vs}, we find a friction coefficient for large $T$ equal to
\begin{equation}
 \mu_\phi=\left(\frac{4\pi  T}{3l}\right)^2\frac{1}{2m\pi\alpha'}+{\cal O}(\omega, l/T).
\end{equation}

\section{Anisotropic drag on a string in 4D Kerr-AdS black hole}

Having recovered the known drag force results in global AdS for any point on the boundary sphere, we are now in a position to introduce anisotropy in the system through angular momentum of the plasma. The corresponding gravitational solution is a Kerr-AdS black hole, and the computation is formally similar to that of the drag force in a charged plasma \cite{Herzog:2006se,Caceres:2006dj,Herzog:2007kh,Atmaja:2010uu}, except that the internal sphere is replaced by the geometric sphere at spatial infinity. This means that we shall need to now consider arbitrary motion of the string on the sphere, rather than special cases. We shall use Kerr-AdS in Boyer-Lindquist coordinates which have less mixing terms than other coordinate system and it manifestly reduces to the non-rotating solution of previous section when the rotation parameter $a$ vanishes. A disadvantage of Boyer-Lindquist coordinates is that this coordinate system does not have manifest $SO(3)$ symmetry far away from the black hole $r \gg M$. Instead one finds AdS in non-standard coordinates \cite{Henneaux:1985tv,Hawking:1998kw,Gibbons:2004ai}. 

Explicitly the metric of four dimensions Kerr-AdS black hole in Boyer-Lindquist coordinates is ~\cite{Hawking:1998kw,Gibbons:2004uw}
\begin{align}
\label{eq50}
ds^2=&\ -\frac{\Delta_r}{\rho^2}\left(dt-\frac{a}{\Xi}\sin^2\theta d\phi\right)^2 +\frac{\rho^2}{\Delta_r}dr^2 +\frac{\rho^2}{\Delta_\theta}d\theta^2\nonumber\\
&\ + \frac{\Delta_\theta\sin^2\theta}{\rho^2}\left(adt-\frac{r^2 +a^2}{\Xi}d\phi\right)^2,
\end{align}
where 
\begin{eqnarray}
\label{eq51}
 \rho^2&=&r^2+a^2\cos^2\theta \nonumber\\
\Delta_r&=&(r^2+a^2)(1+l^2r^2)-2Mr \nonumber\\
\Delta_\theta &=&1-l^2a^2\cos^2\theta \nonumber\\
\Xi &=& 1-l^2a^2,
\end{eqnarray}
with $a$ is a rotation parameter related to the angular momentum $J$ of the black hole: $J=Ma/\Xi^2$. The event horizon or outer horizon is located at $r=r_{KH}$ which is the largest root of $\Delta_r$. The Hawking temperature is given by
\begin{align}
\label{temp}
 T_H=r_{KH}{3l^2r_{KH}^2+ 1+ a^2l^2-{a^2/r_{KH}} \over 4\pi (r_{KH}^2+ a^2)}.
\end{align}
The rotation parameter $a$ is not arbitrary but constrained to $1>a^2l^2$ in order to have a finite positive value of the area~\cite{Hawking:1998kw}.

The action for a string moving in the Kerr-AdS metric is
\begin{eqnarray}
\label{eq52}
S_{NG}&=&-\frac{1}{2\pi\alpha'}\int d\sigma^2\sqrt{-g}, \non
-g&=&\left(\left(a\Delta_r-a(r^2+a^2)\Delta_\theta\right)\frac{\sin^2\theta}{\Xi\rho^2}\phi'+\frac{\rho^2}{\Delta_\theta}\dot{\theta}\theta'
\right.\nonumber
\\ &&
\left.\ \ +\left(\Delta_\theta(r^2+a^2)^2-a^2\Delta_r\sin^2\theta\right)\frac{\sin^2\theta}{\Xi^2\rho^2}\dot{\phi}\phi'\right)^2\notag\\
&&-\left(\frac{\rho^2}{\Delta_r}+\frac{\rho^2}{\Delta_\theta}\theta'^2+\left(\Delta_\theta(r^2+a^2)^2-a^2\Delta_r\sin^2\theta\right)\frac{\sin^2\theta}{\Xi^2\rho^2}\phi'^2\right)\times\notag\\
&&\times\left(\left(a^2\Delta_\theta\sin^2\theta-\Delta_r\right)\frac{1}{\rho^2}+\left(a\Delta_r-a(r^2+a^2)\Delta_\theta\right)\frac{2\sin^2\theta}{\Xi\rho^2}\dot{\phi}\right.\nonumber\\
&&\left.\qquad +\left(\Delta_\theta(r^2+a^2)^2-a^2\Delta_r\sin^2\theta\right)\frac{\sin^2\theta}{\Xi^2\rho^2}\dot{\phi}^2+\frac{\rho^2}{\Delta_\theta}\dot{\theta}^2\right).
\end{eqnarray}
Let us again first consider the equatorial solution for $\theta=\pi/2$. The remaining rotational symmetry around the axis of rotation ensures that $\pi_{\phi}^r$ is still a constant of the motion and substituting the ansatz (\ref{eq7}) the equation of motion now becomes
\begin{eqnarray}
\label{eq55} \eta'(r)&=&\frac{\pi^r_\phi(1-a^2l^2)}{\Delta_r}\sqrt{\frac{(1-a^2l^2-a\omega)^2\Delta_r-f(r)}{\Delta_r-(1-a^2l^2)^2(\pi^r_\phi)^2}},\nonumber\\
f(r)&=&(a-a^3l^2-a^2\omega-\omega r^2)^2,
\end{eqnarray}
Requiring a real solution for $\infty>r>r_{KH}$ demands
  \begin{eqnarray}
 \label{eq:8}
\frac{(1-a^2l^2-a\omega)^2\Delta_r-f(r)}{\Delta_r-(1-a^2l^2)^2(\pi^r_\phi)^2} \geq 0
  \end{eqnarray}
At $r\to\infty$ we recover the relativistic bound 
\begin{equation}
\label{eq57}
 l^2\geq\frac{\omega^2}{(1-a^2l^2-a\omega)^2}.
\end{equation}
for the $r\rar \infty$ boundary Boyer-Lindquist metric 
\begin{eqnarray}
  \label{eq:7}
  ds^2_B = -  \left(\Xi dt- a \sin^2\th d\phi\right)^2  + \frac{1}{l^2\Delta_{\th}\Xi^2} d\th^2 + \frac{\Delta_\th \sin^2\th}{l^2}d\phi^2~. 
\end{eqnarray}
In order to satisfy the inequality (\ref{eq:8}) for other values of $r>r_{KH}$ we need to know the profile of both numerator and denominator on the left hand side of the inequality (\ref{eq:8}). This is not as easy a task as in the case of AdS-Schwarzschild. There could be multiple solutions to the vanishing of the numerator $(1-a^2l^2-a\omega)^2\Delta_r-f(r)$ for $r>r_{KH}$. Following the case of AdS-Schwarzschild we shall insist that at the largest positive root of $(1-a^2l^2-a\omega)^2\Delta_r-f(r)$, denoted $r_{KErg,\ome}$, the numerator and denominator simultaneously change sign as we move from the boundary at $r\to\infty$ toward the horizon $r=r_{KH}$. This fixes $\pi_\phi^r$ to equal
\begin{equation}
 \label{eq58} (\pi^r_\phi)^2= \frac{f(r_{KErg,\ome})}{{(1-a^2l^2)^2(1-a^2l^2-a\omega)^2}}=\frac{(a-a^3l^2-a^2\omega-r_{KErg,\ome}^2\omega)^2}{(1-a^2l^2)^2(1-a^2l^2-a\omega)^2}.
\end{equation}
Then we can compute the drag force as follows
\begin{align}
\label{eq59}
\frac{dp_\phi}{dt}&=-\frac{\pi^r_\phi}{2\pi\alpha'}=-\frac{\sqrt{f(r_{KErg,\ome})}}{2\pi\alpha'(1-a^2l^2-a\omega)(1-a^2l^2)}.
\end{align}
The exact expression for $r_{KErg,\ome}$, as the largest root of $(1-a^2l^2-a\omega)^2\Delta_r(r)-f(r)$ is rather long and complicated and will not be of interest to us. We are primarily interested in the first order correction to the isotropic result, both for experimental reasons and for the more interesting drag force away from the equator. For small $a\ll1$, we can write $r_{KErg,\ome}=r_{SchErg,\ome}+a~r_1+ O(a^2)$, with $r_{SchErg,\ome}$ and
\begin{equation}
 r_1=\frac{\omega l^2 r_{SchErg,\ome}^4-2\omega M r_{SchErg,\ome}}{2(l^2-\omega^2)r_{SchErg,\ome}^3+r_{SchErg,\ome}-M}.
\end{equation}
Then the drag force for a quark moving on the equator of the rotating fluid to first order in $a$ equals
\begin{equation}
\label{conjmom equit}
 \frac{dp_\phi}{dt}=-\frac{1}{2\pi\alpha'}\left[\omega r_{SchErg,\ome}^2-\left(1-\omega r_{SchErg,\ome}\left(\omega r_{SchErg,\ome}+2 r_1\right)\right)a+O(a^2)\right].
\end{equation}
The fist term is the drag force of the 4D non-rotating black hole (\ref{eq13}) and the second term can be considered as correction in the present of small angular momentum $a$. A more illustrative way of writing the drag force is as
\begin{align}
\label{DF a expansion equito}
 \frac{dp_\phi}{dt}&=-\frac{1}{2\pi\alpha'}\left[\pi^0_\phi-a+C~\pi^0_\phi a+O(a^2)\right],\notag\\
C&=\omega\left[\frac{2(2l^2-\omega^2)r_{SchErg,\ome}^3+r_{SchErg,\ome}-5M}{2(l^2-\omega^2)r_{SchErg,\ome}^3+r_{SchErg,\ome}-M}\right],
\end{align}
with $\pi^0_\phi$ is the drag force of non-rotating AdS-Schwarzschild black hole. It is then clear that the first linear term in $a$ is simply due to the relative angular velocities of the black hole and the heavy quark. Choosing a co-rotational velocity such that $\ome r_{SchErg,\ome}^2=a$, there would be no drag to first order as if the quark is at rest with respect to the plasma. On the other hand the term with coefficient $C$ exhibits a nonlinear-enhancement of the drag force in the presence of angular momentum for large velocities $\ome r_{SchErg,\ome} \sim 1$.

\subsection{Static solution}

For equatorial motion the dominant effect of introducing angular momentum is therefore simply a change of frame, as would conform our intuition. Equatorial motion experiences no pressure gradient, however, and as we argued in the introduction this is the effect we wish to extract. To us, the far  interesting situation is to consider how the introduction of angular momentum affects generic non-equatorial great circle motion. True generalization of the arbitrary great circle solutions (\ref{eq32}) to Kerr-AdS black holes is very difficult, however, because of the complexity of the equations of motion. 
Fortunately we shall not need to do so. Our insight builds on the fact that the dominant effect for equatorial motion is simply a change in relative velocity of the quark w.r.t. the plasma. For generic great circle motion, we can solve the equations for the particular situation where we consider the quark to be static, whose precise definition will follow below. This static solution can already capture the drag force effect of a rotating plasma. In or parallel to the equatorial plane, the effect should again be the same as considering a moving string in a non-rotating black hole by switching observers. For motion perpendicular to the equatorial plane, we will find a new component to the force due to the anisotropy breaking by the angular momentum. We expect this new component to be centrifugal force-like and drive the motion away from the poles back to equatorial orbits. Specifically this means that this force will not depend on the direction of the angular momentum, but only on its magnitude. To lowest order in $a$ therefore, this contribution must go as $a^2$.

What we mean precisely by the static solution, is a string solution with the dragging velocity set to zero. Intuitively such a solution, where we keep the quark pinned at one point in the rotating fluid, will give rise to a stationary solution. However, due to the subtlety that the Kerr-AdS metric in Boyer-Lindquist coordinates does not asymptote to a rotation-parameter independent asymptotically AdS metric, this solution is not time-independent in Boyer-Lindquist coordinates. This follows from the coordinate transformation between Boyer-Lindquist and asymptotically AdS coordinates \cite{Hawking:1998kw}
\begin{align}
T&=t,\qquad \Phi=\phi-al^2 t, \notag\\
Y \cos\Theta&= r\cos\theta,\qquad Y^2={1\over \Xi}\left(r^2\Delta_\theta+a^2 \sin^2\theta\right).
\end{align}
Although the full expression for the Kerr-AdS metric in aAdS coordinates $T,Y,\Theta,\Phi$ is very complicated, in the absence of a black hole, for $M=0$, it is simply the global AdS metric
\begin{align}
 ds^2=-(1+l^2Y^2)dT^2+{1\over 1+l^2Y^2}dY^2+Y^2\left(d\Theta^2+\sin^2\Theta d\Phi^2\right).
\end{align}
In the absence of a plasma, the static straight string solution to the equation of motion in this metric , is simply given by $\Theta=\Theta_0$ and $\Phi=\Phi_0$ with $\Phi_0$ and $\Theta_0$ constants corresponding to a massive quark at rest. The corresponding ``time-dependent'' solution in $M=0$ Boyer-Lindquist coordinates is thus
\begin{align}
 \phi&=\Phi_0+a l^2 t, \\
 \theta&=\arccos \left({Y\over r}\cos \Theta_0\right),
\end{align}
with
\begin{align}
 Y^2={r^2(r^2+a^2) \over (1-a^2l^2\sin^2 \Theta_0)r^2+a^2\cos^2 \Theta_0}.
\end{align}
From the derivation it is obvious that this nevertheless describes a static string. 

We now introduce the rotating plasma by considering finite $M$. From previous sections we can see that the straight string with $\ome=0$ is still a solution for a non-rotating plasma. Therefore all bending of the string must be proportional to the rotation parameter $a$. Rather than solving the full complicated equations of motion descending from the action \eqref{eq52}, we can thus try to solve the static string equations order by order in $a$. As we shall see this will already yield non-trivial contributions to the drag force that can be surmised to follow from the rotationally induced pressure gradient. In particular, time-reversal symmetry of the system dictates that the bending in the $\theta$-direction will always be of even order in $a$; to lowest order this is readily verified directly from the equations of motion. 
The force in the $\theta$-direction will be the centrifugal-like force. 
We therefore make the ``static'' ansatz for the curved string solution
\begin{align}
 \theta(r)&=\Theta_0+a^2 \theta_1(r)+{\cal O}(a^4),\\
 \phi(t,r)&=\Phi_0+a l^2 t+ a \phi_1(r)+{\cal O}(a^2).
\end{align}
Solving the equations of motion order by order in power of $a$, we readily obtain
\begin{align}
 \phi_1(r)&=\int dr{P_1\over r^4 h(r)},
\end{align}
with $P_1$ a constant of integration. 
The first correction $\theta_1(r)$ is determined by the inhomogeneous equation
\begin{eqnarray}
  \label{eq:10}
\th_1''+{2\over r^7 h(r)^2}\left(2M^2+r^2+3l^2r^4+2l^4r^6-3Mr-5Ml^2r^3\right)\th_1'& \non
+{1\over 2}l^2r\sin(2\Theta_0)(3l^2r^3-4M)&=&0
\end{eqnarray}
with $h(r)$ as in \eqref{eq1}.
One can solve this with e.g. Mathematica and substituting both first order solutions into the expressions for the conjugate momenta
\begin{eqnarray}
  \label{eq:9}
  \pi^r_{\theta} &=& r^4 h(r) \th_1' a^2+O(a^4) \\\nonumber
\pi^r_{\phi} &=& P_1 \sin(\Theta_0)^2 a+{\cal O}(a^2),
\end{eqnarray}
we find that the world sheet conjugate momenta in radial direction as an expansion of small $a$ near the boundary are given by
\begin{eqnarray}
\pi^r_\theta&= &\left(-3l^2 r+ {2T_2\over\sin(2\Theta_0)}+(1-P_1^2){\log(r)\over M}-{3\over r}+\cdots\right){a^2\over 2}\sin(2\Theta_0)+{\cal O}(a^4),\\\nonumber
  \pi^r_\phi&=& P_1 \sin(\Theta_0)^2 a+{\cal O}(a^2),
\end{eqnarray}
with $T_2$ a constant of integration. Comparing to the equatorial solution ~\eqref{conjmom equit} with $\ome=0$ we see that $P_1=-1$ and $T_2=0$. Because our solution is no longer time-independent, we had to expect a dependence on the radial direction. As is explained in \cite{Herzog:2006gh}, the drag force is read off from the value of $\pi^r_i$ at the AdS boundary. In our case the conjugate momentum $\pi^r_\theta$ diverges linearly as $r\to\infty$ which goes linearly in $r$. This singularity at $r\to\infty$ corresponds to the infinite mass of our heavy quark~\cite{Herzog:2006gh}. In order to have a more realistic picture, we can consider a finite large mass of quark by introducing a cut-off in the geometry near the boundary at $r=r_c$.  In the bulk this is  interpreted as the location of a probe D-brane where the string can end. Following~\cite{Herzog:2006gh}, the static rest mass of quark can be computed. For $a=0,~ T=0$ it is simply $m_{rest}=r_c/{2\pi\alp'}$.
Then by evaluating conjugate momenta above at $r=r_c$ we obtain the leading contribution of the conjugate momenta
\begin{eqnarray}
\pi^r_\theta&=&-\left(6\pi\alpha'l^2m_{rest}\right){a^2\over 2}\sin(2\Theta_0)+\ldots,\\
  \pi^r_\phi&=& -\sin(\Theta_0)^2 a+\ldots,
\end{eqnarray}

\subsection{Drag force}

\begin{figure}[ht]
\begin{minipage}[ht]{0.5\linewidth}
\centering
{\small $a=0.1,M =10$}
  \includegraphics[width=0.3\paperwidth]{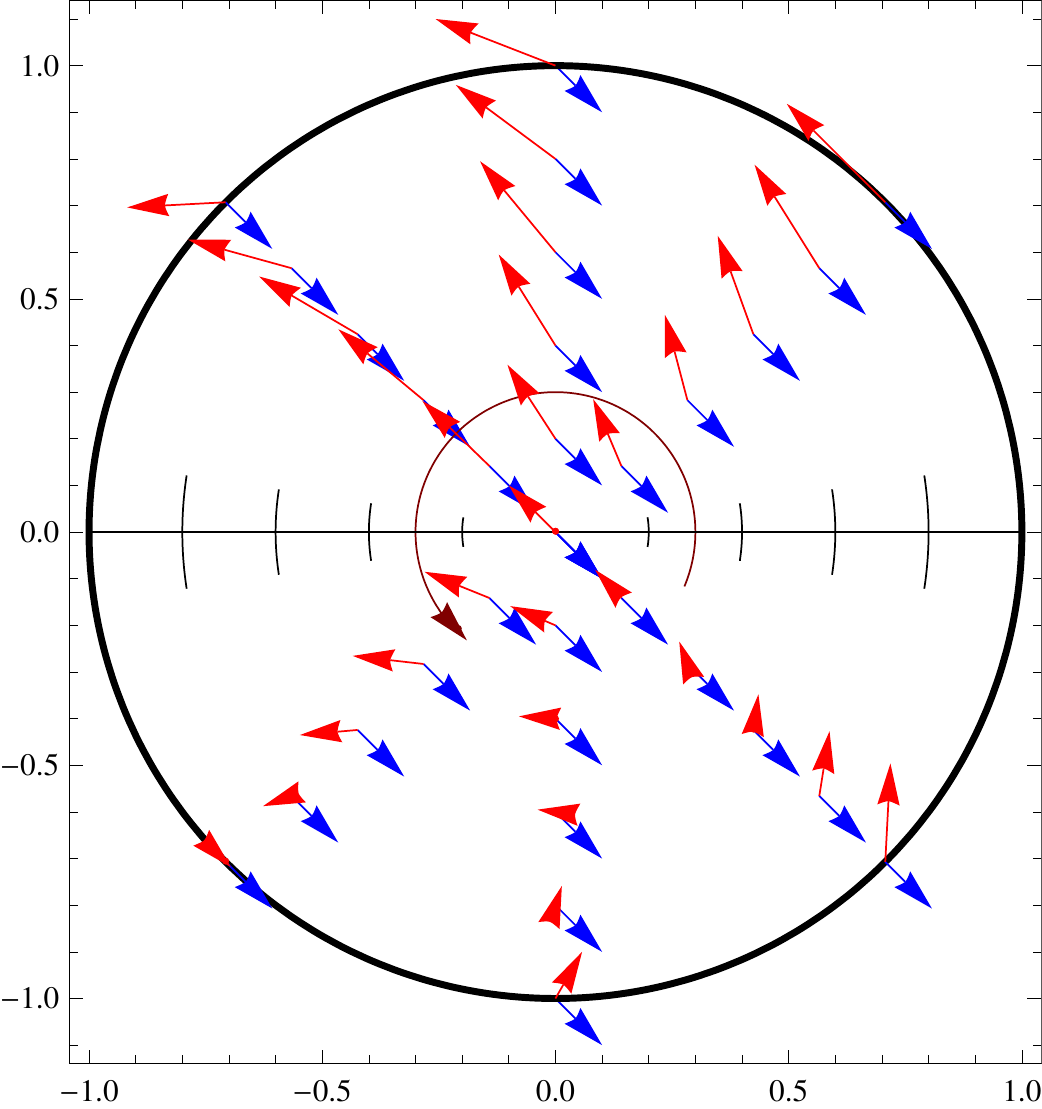}
\end{minipage}
\hspace{0.01\linewidth}
\begin{minipage}[ht]{0.5\linewidth}
\centering
{\small $a=0.1,M=30$}

   \includegraphics[width=0.3\paperwidth]{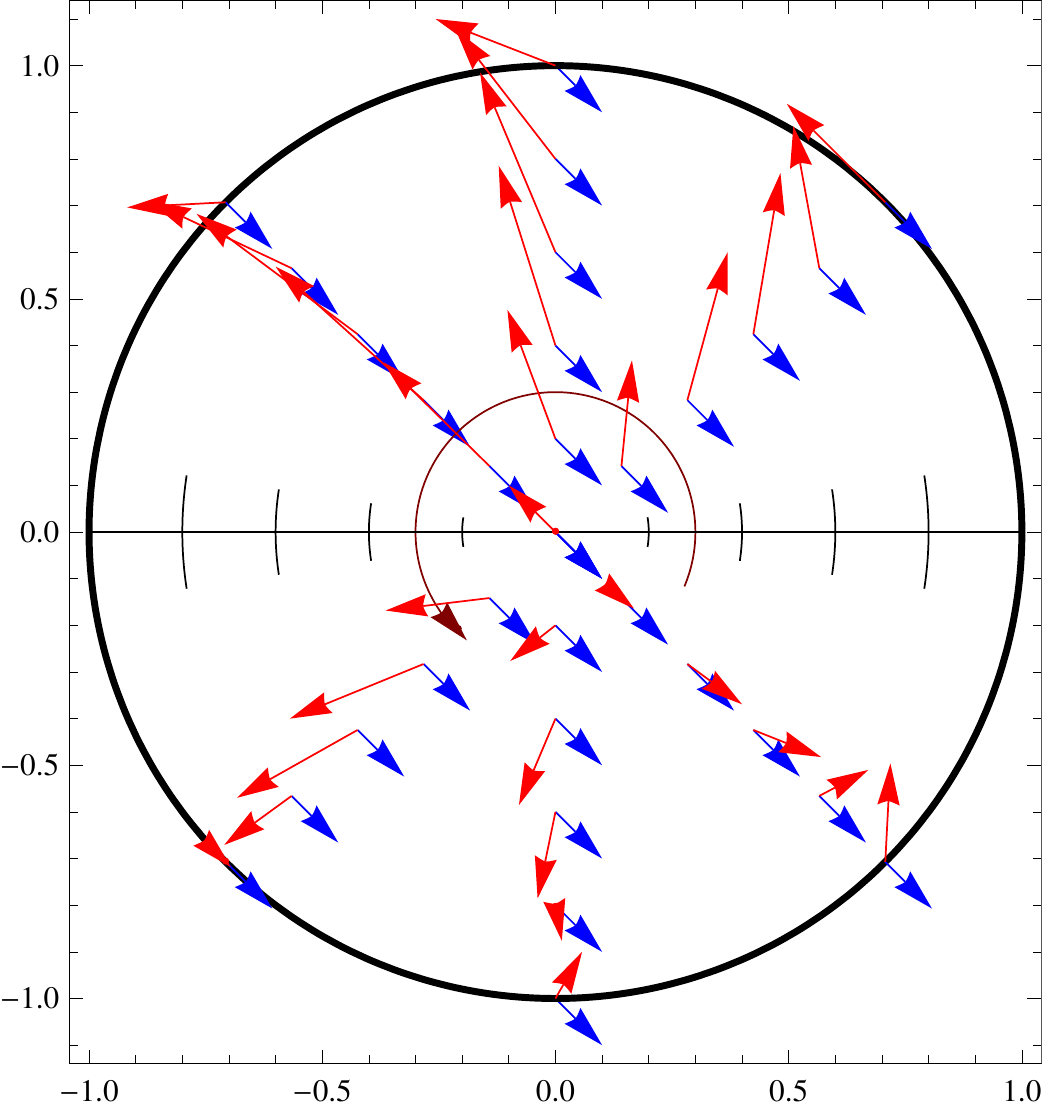}
\end{minipage}
\begin{minipage}[ht]{0.5\linewidth}
\centering
    \includegraphics[width=0.3\paperwidth]{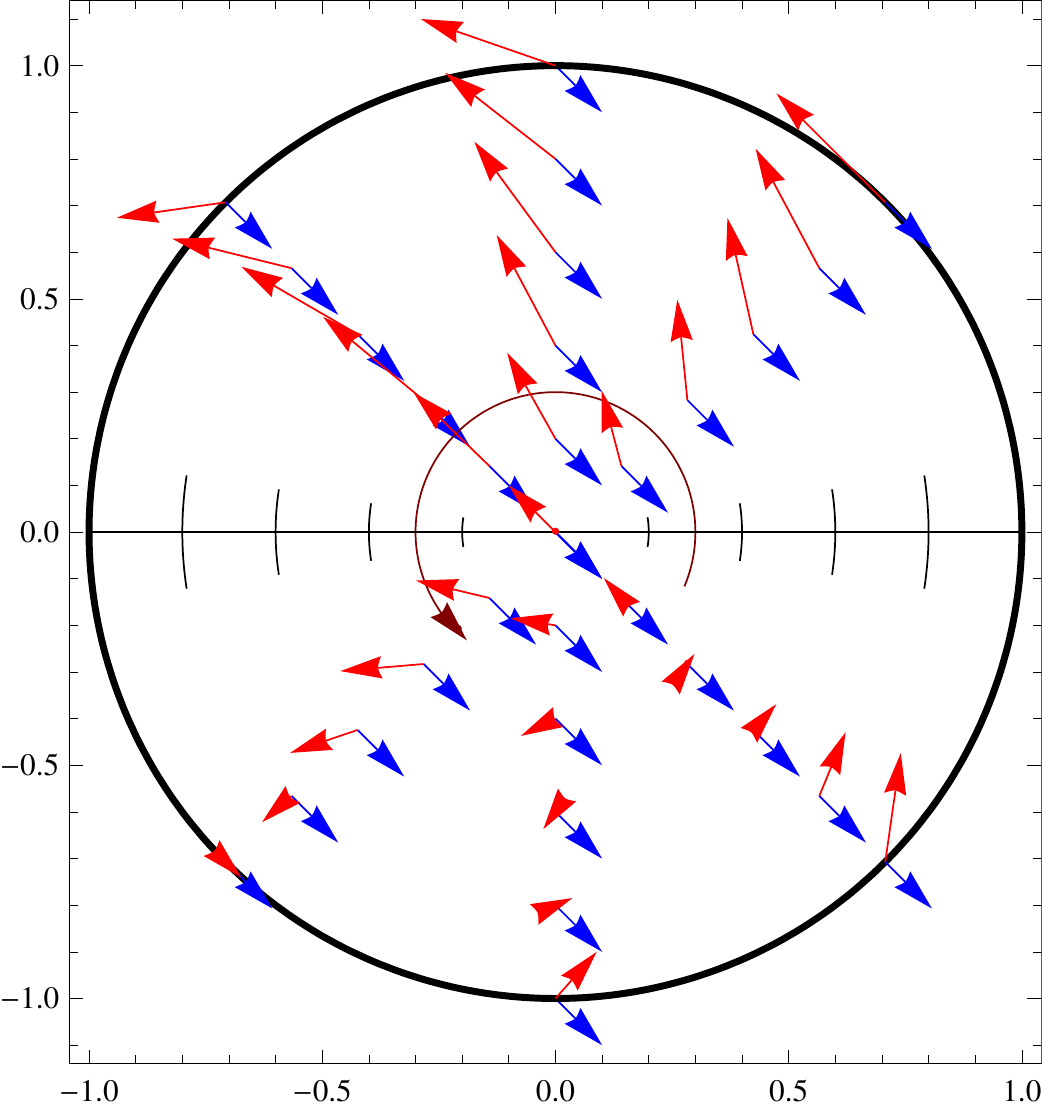}
{\small $a=0.12,M=10$}
\end{minipage}
\hspace{0.01\linewidth}
\begin{minipage}[ht]{0.5\linewidth}
\centering
    \includegraphics[width=0.3\paperwidth]{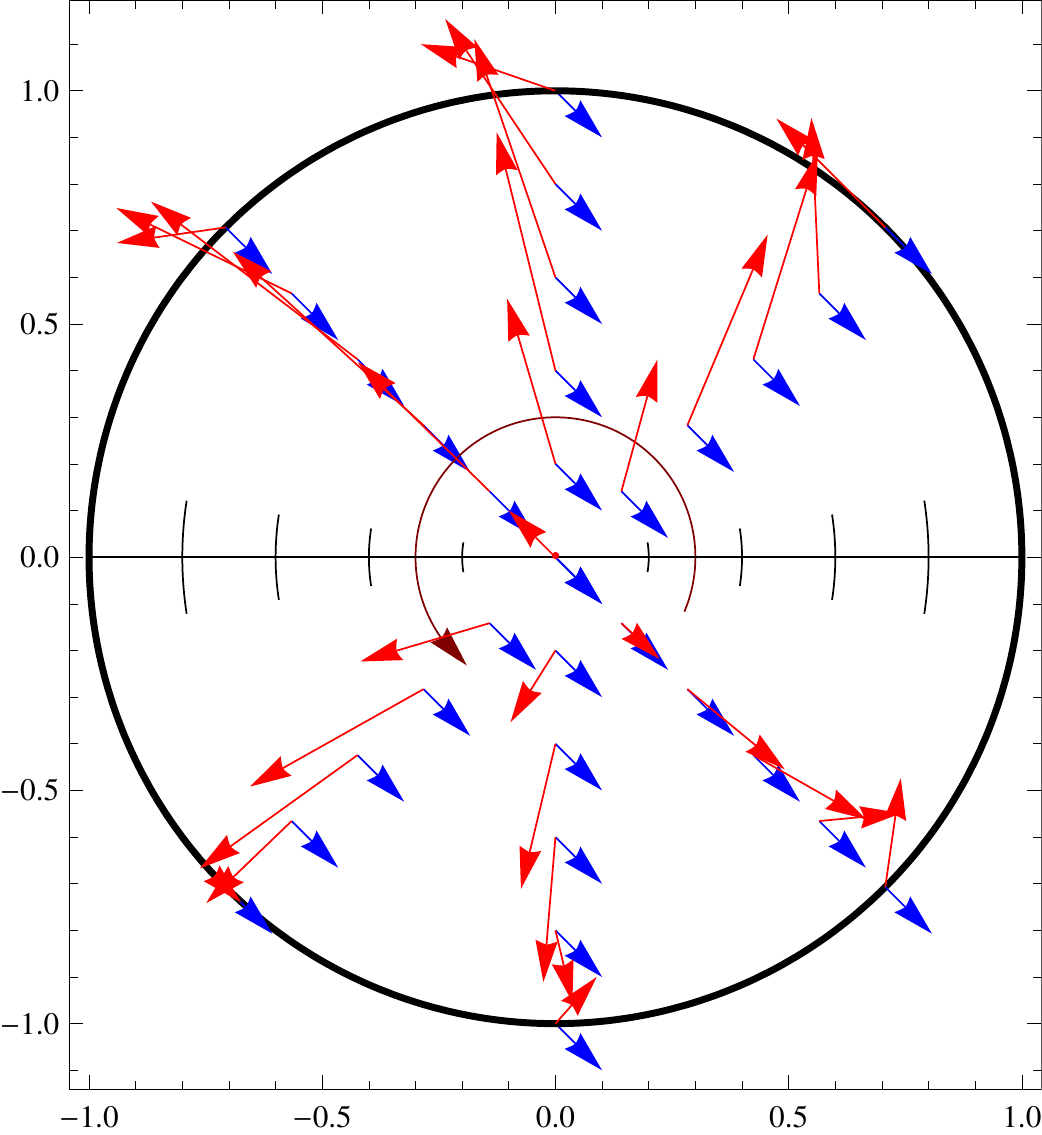}
{\small $a=0.12,M=30$}
\end{minipage}
\caption{ \label{DF1}\small The drag force (red arrow) felt by a heavy quark moving at a uniform velocity (blue arrow) at different positions in a rotating fluid living on a sphere $S^2$ for various rotational velocities $a$  in units of $l^{-1}$ and quark masses $M$ in units of $(6\pi\alp' l)^{-1}$. The sphere is depicted as seen from the North Pole, with the bold line denoting the equator.  The circular brown arrow is the direction of rotation of the fluid. The calibration marks correspond to latitutes $\theta=\left\{\frac{\pi}{10},\frac{2\pi}{10},\frac{3\pi}{10},\frac{4\pi}{10}\right\}$ and the force is in units of $(2\pi\alp'l)^{-1}$.
}
\end{figure}

The conjugate momenta for the static string solution gives us directly the drag force for a static heavy quark held fixed in the rotating plasma. We can, however, combine this result with the known result for the relative velocity-induced drag to obtain an approximation to the general drag force for a slowly moving quark. Since this force vanishes in the absence of rotation, and since the drag force in a non-rotating plasma vanishes for a static quark, to lowest order in both $a$ and $\ome$ the total instantaneous drag force for a quark at location $(\Theta_0,\Phi_0)$ with velocities $(\ome_\th,\ome_\phi)$ is simply the sum of the two individual contributions
\begin{eqnarray}
\label{DF leading order}
  {dp_\theta\over dt}&=&\left(3l^2m_{rest}+{1\over 2\pi\alpha'}\left(2\pi T+\sqrt{4\pi^2 T^2-3l^2}\right)\right){a^2\over 2}\sin(2\Theta_0)-{\omega_\theta r_H^2\over 2\pi\alpha'} +\cO(a,\ome),\non
 {dp_\phi\over dt}&=&-{1\over 2\pi\alpha'}\left(\omega_\phi r_H^2 -\sin(\Theta_0)^2 a\right) +\cO(a,\ome).
\end{eqnarray}
Writing $\sin(2\Theta_0) =2 \sin(\Theta_0)\cos(\Theta_0)$, we clearly see the dragging of the heavy quark towards equatorial motion at $\th=\pi/2$. At the poles, defined at $\Theta_0=0$ and $\Theta_0=\pi$, there is no additional momentum induced drag forces, but these are unstable points. To illustrate the focal aspect of the lowest order instantaneous drag force  for different locations and velocities \eqref{DF leading order} we have give a comprehensive force diagram in \mbox{Fig.~\ref{DF1}}
. The sphere on which the rotating fluid lives is depicted as seen from the north pole and a quark moving at the velocity given by the blue arrow experiences a drag force given by the red arrow.

\section{Discussion and conclusion}

Having computed the leading anisotropic correction to the drag force due to the rotation, we argue that we can extract from this result the leading correction due to general pressure gradients in the system. First recall that any rotating fluid will have a pressure gradient. A relativistic 2+1 dimensional perfect fluid which is rotating at constant angular velocity
 $\Ome$ must live on a two-sphere $S^2$ to be causal. With the metric eq. \eqref{metric boundary} on $S^2\times \RR_t$,
 it has a natural velocity field $u^a=\frac{1}{\sqrt{1-l^{-2}\Omega^2 \sin^2 \th}}(1,0,\Omega)$. In hydrostationary equilibrium its stress-energy tensor
\begin{align}
 T^{ab}=(\rho+P)u^a u^b+P g^{ab},
\end{align}
is conserved, $\nabla_a T^{ab}=0$. Rotational symmetry along $\phi$ dictates that the energy density $\rho$ and pressure $P$ are functions of $\theta$ only. Projecting the conservation equation onto the direction orthogonal to the velocity field
\begin{align}
 (g_{bc} +u_bu_c)\nabla_a T^{ac}=0.
\end{align}
one obtains the non-trivial equation for the pressure
\begin{align}
 \partial_\theta P=\frac{1}{2l^2}{\Omega^2 \sin(2\theta)(P+\rho)\over (1-l^{-2}\Omega^2\sin^2(\theta))} =\frac{1}{2l^2}{\Omega^2 \sin(2\theta) sT\over (1-l^{-2}\Omega^2\sin^2(\theta))}
\end{align}
In the last step we have used the first law of thermodynamics $\rho+P=sT$.

In AdS/CFT the parameters of the fluid are encoded in the parameters of the black hole metric. For rigidly rotating fluids, the mapping has been given in detail in~\cite{Bhattacharyya:2007vs}, and we can identify the angular velocity $\Omega=al^2$.\footnote{The entropy density equals $s=\left(\frac{4\pi T}{(d-1)l}\right)^{d-2} \frac{1}{4G_d(1-a^2)}$ 
; for  AdS${}_4$, dual to $N_c$ M2-branes, Newton's constant equals $G_4=
3l^{-2}/(2N_c)^{3/2}
$; for AdS${}_5$, dual to $N=4$ $SU(N_c)$ SYM, one finds  $G_5= \pi/(2N_c^2l^3)$.
} 
To lowest order in $a$, the pressure gradient in the fluid is therefore
\begin{eqnarray}
  \label{eq:11}
  \pa_{\th} P(\th) = (sT){\frac{a^2l^2}{2} \sin(2\theta)}
\end{eqnarray}

For any constituent of the fluid this pressure gradient provides the centripetal force that ensures rigid rotational motion.  Consider now the situation where the heavy quark is precisely moving with $\ome_{\phi}=a$ and $\ome_{\th}=0$. In that case the quark is at rest w.r.t. the fluid, and the dragforce in the theta-direction is in fact nothing but the ``centrifugal force'' of the particle wishing to persist in its great circle motion. This centrifugal force is equal and opposite to the centripetal force needed if we wanted to keep the heavy quark at fixed value of $\theta$ and co-rotating with the fluid. One way to provide this force is to immerse the heavy quark into a rigidly co-rotating bath of identical quarks; the force then follows from the pressure gradient.  Of course the fluid in AdS, as in real RHIC, already consists of constituents identical to the heavy quark. We therefore put forward that in a general pressure gradient, a heavy quark will experience to first order a force equal and opposite to the $a^2$-component of the dragforce computed in \eqref{DF leading order} where $a^2$ encodes the pressure gradient according to \eqref{eq:11}. If so, we recognize in the resulting expression
\begin{eqnarray}
\label{DF PressGrad}
{dp_\theta\over dt}&=& -\left(3 m_{rest}\right) \frac{\pa_\th P(\th)}{sT} +\ldots
\end{eqnarray}
the standard (relativistic) pressure gradient force (up to the factor $3$), as stated in the introduction. It is interesting to see that AdS/CFT encodes general principles of fluid mechanics once again. 

As a final step, we can subsitute into the expression the entropy density of the strongly coupled gauge theory as computed in AdS/CFT. In the specific case of the strongly coupled $d=3$ ABJM/M2-brane theory, we find
\begin{eqnarray}
  \label{eq:13}
  {dp_\theta\over dt}&=& -\left(3 m_{rest}\right) \frac{3^3}{4\pi^2 (2N_c)^{3/2}T^3}{\pa_\th P(\th)}
\end{eqnarray}
At the same time, the generality of the pressure gradient force implies that the result \eqref{DF PressGrad} is the same in any dimension (up to the numerical factor 3). On the gravity side, one can explain this by realizing that in any dimension the rotational dependence of a simple rotating black hole is characterized by a single non-zero element  of the Cartan algebra of the rotation group. Extrapolating the result to the more relevant case of $d=4$ YM and using the entropy density for strongly coupled $N=4$ SYM, the anisotropic component of the drag force equals
\begin{eqnarray}
  \label{eq:14}
  {dp_\theta\over dt}&=& -\left(3 m_{rest}\right) \frac{2}{\pi^2N_c^2T^4}{\pa_\th P(\th)}.
\end{eqnarray}

\acknowledgments
We are grateful to P. McFadden, T. Peitzmann, S. Ross, R. Snellings, M. Taylor, J. Hoogeveen, and U. Wiedemann for very helpful discussions. KS thanks the Galileo Galilei Institute for Theoretical Physics for the hospitality and the INFN for partial support during the completion of this work. This research was supported in part by a VIDI Innovative Research Incentive Grant from the Netherlands Organisation for Scientific Research (NWO) and a Programme Grant from the Dutch Foundation for Fundamental Research on Matter (FOM).

\end{document}